# A variational quantum algorithm-based numerical method for solving potential and Stokes flows

Y. Y. Liu[a], Z. Chen[b], C. Shu[a,*], P. Rebentrost[c], Y. G. Liu[a], S. C. Chew[d], B. C. Khoo[a] and Y. D. Cui[d]

[a] Department of Mechanical Engineering, National University of Singapore, 10 Kent Ridge Crescent, Singapore 119260

[b] School of Naval Architecture, Ocean and Civil Engineering, Shanghai Jiao Tong University, Shanghai, China 200240

[c] Center for Quantum Technologies, National University of Singapore, 3 Science Drive 2, Singapore 117543, Singapore

[d] National University of Singapore, Singapore 117411, Republic of Singapore

## Abstract

This paper presents a numerical method based on the variational quantum algorithm to solve potential and Stokes flow problems. In this method, the governing equations for potential and Stokes flows can be respectively written in the form of Laplace's equation and Stokes equations using velocity potential, stream function and vorticity formulations. Then the finite difference method and the generalised differential quadrature (GDQ) method are applied to discretize the governing equations. For the prescribed boundary conditions, the corresponding linear systems of equations can be obtained. These linear systems are solved by using the variational quantum linear solver (VQLS), which resolves the potential and Stokes flow problems equivalently. To the best of authors' knowledge, this is the first study that incorporates the GDQ method which is inherently a high-order discretization method with the VQLS algorithm. Since the GDQ method can utilize much fewer grid points than the finite difference method to approximate derivatives with a higher order of accuracy, the size of the input matrix for the VQLS algorithm can be smaller. In this way, the computational cost may be saved. The performance of the present method is comprehensively assessed by two representative examples, namely, the potential flow around a circular cylinder and Stokes flow in a lid-driven cavity. Numerical results validate the applicability and accuracy of the present VQLS-based method. Furthermore, its time complexity is evaluated by the heuristic scaling, which demonstrates that the present method scales efficiently in the number of qubits $n$ and the precision $\varepsilon$. This work brings quantum computing to the field of computational fluid dynamics. By virtue of quantum advantage over classical methods, promising advances in solving large-scale fluid mechanics problems of engineering interest may be prompted.

---

* Corresponding author, E-mail: mpeshuc@nus.edu.sg (C. Shu).

# 1. Introduction

Potential flow and Stokes flow are two cornerstones of fluid dynamics. As an idealized model of fluid flow, potential flow is characterized by an inviscid and irrotational flow field in which the velocity vector can be expressed as the gradient of the velocity potential function. In the case of incompressible flows, the velocity potential satisfies Laplace's equation. Stokes flow, known as creeping flow synonymously, is a particular type of fluid flow which assumes that the inertia force is negligible compared with the viscous and pressure forces. Normally, for Stokes flows, the fluid velocities are extremely slow and the physical viscosities are very large, which can be characterized by a very small Reynolds number. The motion of Stokes flow is described by the Stokes equations. In fact, both Laplace's equation for potential flows and Stokes equations for Stokes flows are simplified from Navier-Stokes (N-S) equations with specific assumptions such as zero vorticity and zero viscosity for Laplace's equation and very low Reynolds number ($Re \ll 1$) for Stokes equations. These simplified equations make many fluid problems resolvable in the general case and provide good practice guidance for analysing flow problems in physics, engineering and biology. Practical applications of potential flow theory can be found in studies of aircraft [1,2], ship motion [3,4], water waves [5-8], and groundwater flow [9]. Stokes flow is widely employed in investigations of natural phenomena [10-13] such as microorganisms swimming and fluid flow through small cracks, and in engineering applications [14-16] such as paint, micro-electro-mechanical systems and flexible filaments/fibers immersed in a fluid.

The Laplace's equation and Stokes equations can be solved analytically and numerically. The analytical approaches aim to retrieve the exact solutions, while the numerical methods approach to the approximate solutions. The latter refer to the techniques of computational fluid dynamics (CFD) which use numerical algorithms to compute solutions on computers. In past decades, numerical studies using CFD have revolutionised the investigations of the potential flows and Stokes flows as well as related fluid mechanics problems. However, the slowdown of the rising rate in the computing capabilities of classical computers is being perceived and it may be durable unless a revolution in hardware design occur [17,18]. Fortunately, the advent and breakthrough of quantum computing (QC) on quantum devices provide another paradigm in scientific computing. Unlike classical computing techniques, QC conducts computations by harnessing the laws of quantum mechanics, which is expected to be more competitive in solving complex problems [19-21]. For instance, the Harrow-Hassidim-Lloyd (HHL) algorithm [22] was reported to have an exponential speedup compared with classical methods in solving some types of linear systems of equations.

The promising computing power of QC stimulates the construction of numerical algorithms for CFD which are deployable on quantum computers. Typical works involving practical implementations include HHL based studies [23-27], variational quantum algorithms (VQAs) based studies [28-31] and quantum machine learning (QML) based studies [32-34]. Specifically, in research based on the HHL algorithm,

HHL serves as a subroutine to solve the linear systems formed by discretizing the target partial differential equations (PDEs) with numerical methods. Although applications to resolve some flow problems have been presented, the expected exponential speedup of HHL algorithm has not been clearly demonstrated. In addition, since the HHL-type algorithms are based on the Quantum Fourier Transform (QFT) and thus require large circuit resources, they may be not applicable to practical flow problems on the current Noisy Intermediate Scale Quantum (NISQ) [35] devices.

As recently developed quantum algorithms, the VQAs integrate the QC with classical computing technologies, which seems to be more applicable on the NISQ hardware. In practice, one successful application of VQAs to solve linear systems is the variational quantum linear solver (VQLS) [36-39]. It has been applied to acquire numerical solutions to PDEs such as the Poisson equation [31,40,41] and the heat conduction equation [30]. Nevertheless, practical applications of the VQLS algorithm to flow problems in engineering contexts remain scarce.

Based on the aforementioned discussions, this work will explore the application of the VQLS-based algorithm [30] to resolve the potential flow and the Stokes flow problems with engineering significance. The Laplace's equation and the Stokes equations will be discretized by the finite difference and generalized differential quadrature (GDQ) [42] methods with given boundary conditions. Then the resulting linear systems of equations will be solved by the VQLS algorithm. Note that, different from the finite difference method, the GDQ method is inherently a high-order discretization method which can utilize less grid points for an expected accuracy. As a result, smaller sizes of input matrices for the VQLS algorithm can be expected and so is the reduced computational cost. The importance of potential flow and Stokes flow which correspond closely to real-life flows over the whole of fluid mechanics highlight the necessity of this study. In addition, many valuable insights arising from this study may lead to the extensive application of quantum computing to more complicated fluid flow problems.

The rest of this paper is organized as follows. Section 2 details the proposed VQLS-based direct numerical method for simulating potential and Stokes flows. In specific, governing equations of potential and Stokes flows are first revisited. Numerical discretization for target governing equations and technical details of the VQLS algorithm are then presented. Section 3 illustrates practical applications of the present method. Finally, Section 4 concludes this work.

## 2. Variational Quantum Linear Solver based Direct Numerical Method for Simulation of Potential and Stokes Flows

Considering the significance of potential and Stokes flows and motivated by the promising advantage of quantum computing, this section is devoted to introduction of a hybrid VQLS-based direct numerical method for effectively simulating potential and Stokes flow. First, governing equations for these two flows are revisited. Then, the finite difference method and the generalized differential quadrature method are introduced to discretize the governing equations with appropriate boundary conditions. As a result, the corresponding linear systems of equations are obtained. These linear systems are solved by the VQLS algorithm, which equivalently provides solutions for potential and Stokes flows.

## 2.1. Governing equations

### 2.1.1 Potential flow

In the case of a two-dimensional incompressible potential flow, the velocity potential $\varphi$ with the definition of $\boldsymbol{u} = \nabla \varphi$ satisfies the following Laplace's equation in Cartesian coordinates.

$$\nabla \cdot \boldsymbol{u} = \nabla^2 \varphi = \frac{\partial^2 \varphi}{\partial x^2} + \frac{\partial^2 \varphi}{\partial y^2} = 0, \qquad (1)$$

where $\boldsymbol{u}$ denotes the velocity vector. This Laplace's equation also can be written in cylindrical or polar coordinates by using the following conversions.

$$\begin{aligned} x &= r\cos(\theta), \\ y &= r\sin(\theta), \end{aligned} \qquad (2)$$

where $r$ denotes the radius and $\theta$ represents the azimuth angle. Accordingly, the Laplace's equation in Eq. (1) takes the form as follows,

$$\frac{\partial^2 \varphi}{\partial r^2} + \frac{1}{r}\frac{\partial \varphi}{\partial r} + \frac{1}{r^2}\frac{\partial^2 \varphi}{\partial \theta^2} = 0. \qquad (3)$$

### 2.2.1 Stokes flow

The vorticity-stream function method is one popular approach for solving two-dimensional incompressible flows. By introducing the vorticity $\omega$ and the stream function $\psi$, the vorticity-stream function form of N-S equations can be presented in Cartesian coordinates as follows.

$$\begin{cases} \text{Re}\left(\dfrac{\partial \omega}{\partial t} + \dfrac{\partial \psi}{\partial y}\dfrac{\partial \omega}{\partial x} - \dfrac{\partial \psi}{\partial x}\dfrac{\partial \omega}{\partial y}\right) = \dfrac{\partial^2 \omega}{\partial x^2} + \dfrac{\partial^2 \omega}{\partial y^2}, \\ \nabla^2 \psi = \dfrac{\partial^2 \psi}{\partial x^2} + \dfrac{\partial^2 \psi}{\partial y^2} = -\omega. \end{cases} \quad (4)$$

Under Stokes flow conditions, the Reynolds number Re is close to zero, which yields the following equations from Eq. (4).

$$\begin{cases} 0 = \dfrac{\partial^2 \omega}{\partial x^2} + \dfrac{\partial^2 \omega}{\partial y^2}, \\ \dfrac{\partial^2 \psi}{\partial x^2} + \dfrac{\partial^2 \psi}{\partial y^2} = -\omega. \end{cases} \quad (5)$$

To directly compute the stream functions $\psi$, Eq. (5) is further rewritten as follows.

$$\nabla^4 \psi = \dfrac{\partial^4 \psi}{\partial x^4} + \dfrac{\partial^4 \psi}{\partial y^4} + 2\dfrac{\partial^4 \psi}{\partial x^2 \partial y^2} = 0, \quad (6)$$

which is a biharmonic equation.

## 2.2. Generation of linear systems of equations

### 2.2.1 Potential flow

In both mathematics and fluid dynamics, potential flow around a circular cylinder is a classical problem which provides foundations for further investigations on other problems with more complex geometries such as the airfoil, the ship hull and propellers. For convenience, Eq. (3) for potential flow is considered in this work to compute the solution for the velocity potential $\varphi$ in polar coordinates. With the discrete values being denoted by $\varphi_{i,j} = \varphi(r_i, \theta_j)$, discretizing Eq. (3) with the central difference formula, for $i = 2, 3, \ldots, N\text{-}1, j = 1, 2, \ldots, M\text{-}1$, gives

$$\frac{\varphi_{i+1,j} - 2\varphi_{i,j} + \varphi_{i-1,j}}{\Delta r^2} + \frac{1}{r_i}\frac{\varphi_{i+1,j} - \varphi_{i-1,j}}{2\Delta r} + \frac{1}{r_i^2}\frac{\varphi_{i,j+1} - 2\varphi_{i,j} + \varphi_{i,j-1}}{\Delta \theta^2} = 0, \qquad (7)$$

where $\Delta r = (R_{out} - R_{in})/(N-1)$, $\Delta \theta = 2\pi/(M-1)$ and $r_i = R_{in} + (i-1)\Delta r$ with $R_{out}$, $R_{in}$, $N$ and $M$ being the outer radius of the mesh, the radius of the cylinder, the numbers of grid points in $i$ and $j$ directions, respectively. By introducing $a_i = \Delta r/2r_i$ and $b_i = \Delta r^2/(r_i^2 \Delta \theta^2)$, Eq. (7) can be rewritten as follows

$$\varphi_{i+1,j}(1+a_i) + \varphi_{i-1,j}(1-a_i) + \varphi_{i,j+1}b_i + \varphi_{i,j-1}b_i - \varphi_{i,j}(2+2b_i) = 0. \qquad (8)$$

Given the free-stream velocity $V_0$, the practical implementation includes the following conditions on boundaries of the computational domain.

$$\begin{cases} \varphi_{1,j} = c, & \text{cylinder surface,} \\ \varphi_{N,j} = V_0 y_{N,j} + c, & \text{outer boundary,} \\ \varphi_{i,0} = \varphi_{i,M-1}, \varphi_{i,1} = \varphi_{i,M}, & \text{periodic boundary conditon at } y = 0, \end{cases} \qquad (9)$$

where is $c$ a constant. Solving such a problem is equivalent to solving systems of linear equations as follows,

$$A\vec{x} = \vec{b}, \qquad (10)$$

with

$$\vec{x} = \begin{bmatrix} \varphi_{1,1} & \varphi_{2,1} & \cdots & \varphi_{N,1} & \varphi_{1,2} & \varphi_{2,2} & \cdots & \varphi_{N,2} & \cdots & \varphi_{N-1,M} & \varphi_{N,M} \end{bmatrix}^T_{NM \times 1},$$

$$\vec{b} = \begin{bmatrix} c & 0 & \cdots & 0 & c+V_0 y_{N,1} & c & 0 & \cdots & 0 & c+V_0 y_{N,2} & \cdots & c & 0 & \cdots & 0 & c+V_0 y_{N,1} \end{bmatrix}^T_{NM \times 1}.$$

For convenience, the unknowns $\varphi_{i,j}$ are ordered by first grouping the grid points along the same azimuth angle, then moving counterclockwise to cover the whole domain. Accordingly, the vector $\vec{\phi}$ is introduced with the following definition to replace $\vec{x}$.

$$\vec{\phi} = \begin{bmatrix} \vec{\varphi}_1 \\ \vec{\varphi}_2 \\ \vdots \\ \vec{\varphi}_M \end{bmatrix}, \quad \vec{\varphi}_j = \begin{bmatrix} \varphi_{1,j} \\ \varphi_{2,j} \\ \vdots \\ \varphi_{N,j} \end{bmatrix}. \tag{11}$$

The remaining problem is to solve a large sparse linear system $A\vec{\phi} = \vec{b}$, where the matrix $A$ is

$$A = \begin{bmatrix} T-2D & D & & & D & 0 \\ D & T-2D & D & & & \\ & \ddots & \ddots & \ddots & & \\ & & \ddots & \ddots & \ddots & \\ & & & \ddots & \ddots & \ddots \\ & & & D & T-2D & D \\ -I & & & & 0 & I \end{bmatrix}_{NM \times NM}, \tag{12}$$

with a diagonal matrix $D = diag(0, b_2, b_3, \ldots b_{N-1}, 0)$ and

$$T = \begin{bmatrix} 1 & 0 & & & & \\ 1-a_2 & -2 & 1+a_2 & & & \\ & \ddots & \ddots & \ddots & & \\ & & \ddots & \ddots & \ddots & \\ & & & \ddots & \ddots & \ddots \\ & & & 1-a_{N-1} & -2 & 1+a_{N-1} \\ & & & & 0 & 1 \end{bmatrix}_{N \times N} \tag{13}$$

### 2.2.2 Stokes flow

For illustrative purposes, Stokes flow in a lid-driven cavity which is an important benchmark is studied here as a representative example. Its boundary conditions are illustrated in Fig. 1. The governing equation (6) in Cartesian coordinates is discretized by two numerical discretization methods, namely, the central finite difference method and the differential quadrature method.

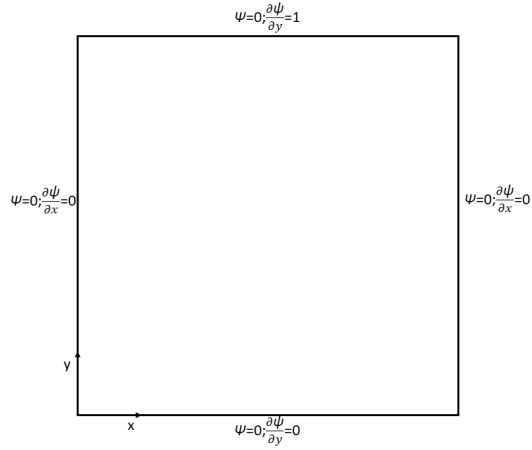

Fig. 1. Schematic diagram of lid-driven cavity flow problem.

A. Discretization with finite difference method

First, after discretization with the central difference formula on a uniform mesh of $N \times N$ grid points, the discretized form can be expressed as follows for the interior domain, i.e., $i = 2, 3, \ldots, N\text{-}1, j = 2, 3, \ldots, N\text{-}1$.

$$\frac{\psi_{i+2,j} - 4\psi_{i+1,j} + 6\psi_{i,j} - 4\psi_{i-1,j} + \psi_{i-2,j}}{h^4} + \frac{\psi_{i,j+2} - 4\psi_{i,j+1} + 6\psi_{i,j} - 4\psi_{i,j-1} + \psi_{i,j-2}}{h^4} + \frac{\left[\psi_{i+1,j+1} - 2\psi_{i,j+1} + \psi_{i-1,j+1} - 2\left(\psi_{i+1,j} - 2\psi_{i,j} + \psi_{i-1,j}\right) + \psi_{i+1,j-1} - 2\psi_{i,j-1} + \psi_{i-1,j-1}\right]}{h^4} = 0,$$

(14)

where $\psi_{i,j}$ is the stream function at a mesh point $(x_i, y_j)$ and $h$ denotes the mesh size. Then we have

$$\psi_{i+2,j} + \psi_{i-2,j} + \psi_{i,j+2} + \psi_{i,j-2} + \psi_{i+1,j+1} + \psi_{i-1,j+1} + \psi_{i+1,j-1} + \psi_{i-1,j-1} \\ -6\left(\psi_{i+1,j} + \psi_{i-1,j} + \psi_{i,j+1} + \psi_{i,j-1}\right) + 16\psi_{i,j} = 0.$$

(15)

As illustrated in Fig. 1, the boundary conditions are given as $\psi = 0$ and the gradients normal to the boundary $\psi_n = 1$ on the top boundary, and $\psi = \psi_n = 0$ on the other three sides. Clearly, $\psi_n$ is discontinuous at the top corners and thus $\psi$ has singularities there. The bottom corners exist weaker singularities associated with the formation of Moffatt eddies whereas $\psi_n$ is continuous. In the literature, there are several methods [43-45] to well address such a problem. Here, in a simplified way, the boundary conditions are enforced via the following relations.

$$\begin{cases} \psi_{1,j} = \psi_{N,j} = \psi_{i,1} = \psi_{i,N} = 0, \\ \psi_{0,j} = \psi_{2,j}, \psi_{N+1,j} = \psi_{N-1,j}, \psi_{i,0} = \psi_{i,2}, \quad i=1,...,N, j=1,...,N-1. \\ \psi_{i,N+1} = \psi_{i,N-1} - 2h, \end{cases} \quad (16)$$

As a result, the solution of Eq. (15) can be equivalently computed from solving a system of linear equations $A\vec{x} = \vec{b}$ with the unknown vector $\vec{x}$ and the vector $\vec{b}$ being

$$\vec{x} = \begin{bmatrix} \psi_{2,2} & \psi_{3,2} & \cdots & \psi_{N-1,2} & \psi_{2,3} & \psi_{3,3} & \cdots & \psi_{N-1,3} & \cdots & \psi_{N-2,N-1} & \psi_{N-1,N-1} \end{bmatrix}^T_{(N-2)^2 \times 1},$$

$$\vec{b} = \begin{bmatrix} 0 & 0 & \cdots & 0 & 0 & 0 & 0 & \cdots & 0 & 0 & \cdots & 0 & 2 & \cdots & 2 & 2 \end{bmatrix}^T_{(N-2)^2 \times 1}.$$

Similar to the case of potential flow, the unknowns $\psi_{i,j}$ can be computed through $\vec{\phi}$ which is defined by

$$\vec{\phi} = \begin{bmatrix} \vec{\psi}_2 \\ \vec{\psi}_3 \\ \vdots \\ \vec{\psi}_{N-1} \end{bmatrix}, \quad \vec{\psi}_j = \begin{bmatrix} \psi_{2,j} \\ \psi_{3,j} \\ \vdots \\ \psi_{N-1,j} \end{bmatrix}. \quad (17)$$

Thus, the linear system to be solved is

$$A\vec{\phi} = \vec{b}, \quad (18)$$

where $A$ can be written as

$$A = \begin{bmatrix} T+D+I & G & I & & & & & \\ G & T+D & \ddots & \ddots & & & & \\ I & \ddots & \ddots & \ddots & \ddots & & & \\ & \ddots & \ddots & \ddots & \ddots & \ddots & & \\ & & \ddots & \ddots & \ddots & \ddots & & I \\ & & & \ddots & \ddots & T+D & G \\ & & & & I & G & T+D+I \end{bmatrix}_{(N-2)^2 \times (N-2)^2} \quad (19)$$

with $D = diag(1,0,\ldots 0,1)$ and

$$T = \begin{bmatrix} 16 & -6 & 1 & & & & \\ -6 & 16 & \ddots & \ddots & & & \\ 1 & \ddots & \ddots & \ddots & \ddots & & \\ & \ddots & \ddots & \ddots & \ddots & \ddots & \\ & & \ddots & \ddots & \ddots & \ddots & 1 \\ & & & \ddots & \ddots & 16 & -6 \\ & & & & 1 & -6 & 16 \end{bmatrix}_{(N-2)\times(N-2)} \quad (20)$$

$$G = \begin{bmatrix} -6 & 1 & & & & \\ 1 & \ddots & \ddots & & & \\ & \ddots & \ddots & \ddots & & \\ & & \ddots & \ddots & \ddots & \\ & & & \ddots & \ddots & 1 \\ & & & & 1 & -6 \end{bmatrix}_{(N-2)\times(N-2)} \quad (21)$$

B. Discretization with generalized differential quadrature

Different from the central difference method used in the preceding discretization which is only second-order accurate, the GDQ method commonly serves as a high-order numerical discretization technique for derivative approximation. Its idea originates from the conventional integral quadrature, where any integral over a closed domain can be approximated by a linear weighted sum of functional values at all points in the integral domain. Simply, the GDQ method linearly combines the functional values at all discrete points along a mesh line to approximate partial derivatives in that direction. To be specific, three fourth-order partial derivatives of the stream function $\psi_{i,j}$ at an interior mesh point ($x_i, y_j$) in Eq. (6) can be approximated by the GDQ method as follows:

$$\frac{\partial^4 \psi}{\partial x^4} = \sum_{k=1}^{N} w_{i,k}^{(4)} \psi_{k,j},$$

$$\frac{\partial^4 \psi}{\partial y^4} = \sum_{k=1}^{N} \overline{w}_{j,k}^{(4)} \psi_{i,k}, \qquad (22)$$

$$\frac{\partial^4 \psi}{\partial x^2 \partial y^2} = \frac{\partial^2}{\partial x^2}\left(\frac{\partial^2 \psi}{\partial y^2}\right) = \frac{\partial^2}{\partial y^2}\left(\frac{\partial^2 \psi}{\partial x^2}\right) = \sum_{l=1}^{N}\sum_{k=1}^{N} w_{i,l}^{(2)} \overline{w}_{j,k}^{(2)} \psi_{l,k},$$

where $w_{i,j}^{(m)}$ and $\overline{w}_{i,j}^{(m)}$ denote the weighting coefficients for the $m$th-order partial derivatives in $x$ and $y$ directions, respectively. In a similar manner, the Neumann boundary conditions shown in Fig. 1 can be implemented with the Dirichlet boundary conditions following the approach in Ref. [46].

$$\frac{\partial \psi}{\partial x} = \sum_{k=1}^{N} w_{i,k}^{(1)} \psi_{k,j} = 0, \quad \text{for } i = 1, N, j = 2, 3, \ldots, N-1,$$

$$\left.\frac{\partial \psi}{\partial y}\right|_{j=1} = \sum_{k=1}^{N} \overline{w}_{1,k}^{(1)} \psi_{i,k} = 0, \left.\frac{\partial \psi}{\partial y}\right|_{j=N} = \sum_{k=1}^{N} \overline{w}_{N,k}^{(1)} \psi_{i,k} = 1, \quad \text{for } i = 2, 3, \ldots, N-1. \qquad (23)$$

Given the grid distribution in each spatial dimension, the unknown coefficients $w_{i,j}^{(m)}$ and $\overline{w}_{i,j}^{(m)}$ in above equations can be explicitly and easily computed through a simple algebraic formulation and a recurrence relationship [42,46,47]. These weighting coefficients comprise a sparse matrix $A$ and the unknown stream function can be computed by solving the linear system of equations $A\vec{\psi} = \vec{b}$. Fig. 2 plots the regular sparsity patterns of matrices $A$ when $N$ = 6, 8 and 10, where the dots denote the nonzero entries.

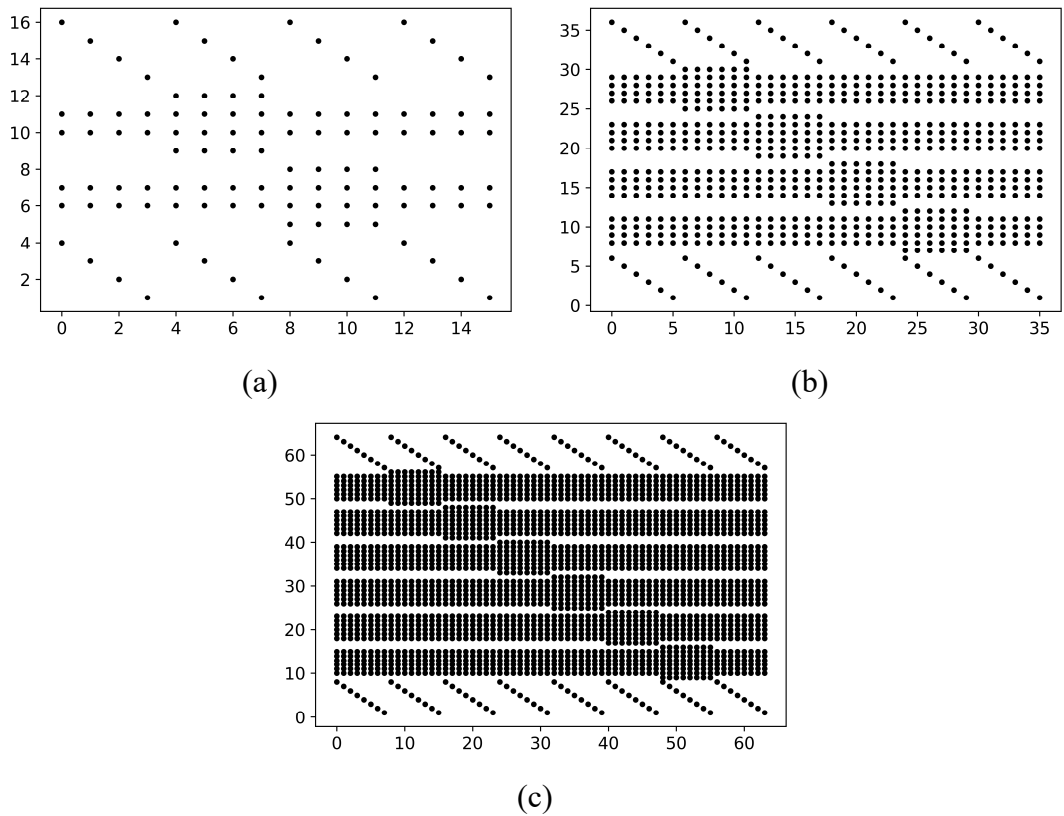

Fig. 2. The regular sparsity patterns of matrices $A$ when (a) $N = 6$, (b) $N = 8$ and (c) $N = 10$. Entries with nonzero values are represented by a black dot.

### 2.3. Solution to linear systems of equations using variational quantum linear solver

Solutions of the above linear systems are solved by the VQLS algorithm, which directly resolves the potential and Stokes flow problems. The VQLS is a variational quantum algorithm for solving linear systems of equations on NISQ quantum computers. The method may exhibit advantages for certain large liner systems which can be implemented efficiently on the quantum device. In practical implementation, for the linear systems generated previously in Subsection 2.2, the VQLS can find a normalized $|x\rangle$ satisfying the relationship $A|x\rangle=|b\rangle$ where $|b\rangle$ is the quantum state prepared from the known vector $\vec{b}$. The inputs to the VQLS algorithm are the matrix $A$ which is a linear combination of unitary matrices $A_m$ with the coefficients $c_m$ and a short-depth quantum circuit $U$ with $|b\rangle=U|0\rangle$ (see also Fig. 3). After devising a parameterized ansatz $V(\alpha)$, a cost function $C(\alpha)$ is constructed and evaluated in the hybrid quantum-classical optimization loop. If the convergence criterion for the cost function $C(\alpha)$ is achieved, the optimal parameters $\alpha^*$ for the ansatz circuit are ready to prepare outputs. The final output is the quantum state $|x^*\rangle=V(\alpha^*)|0\rangle$ that is proportional to the solution vector $\vec{x}$. In this subsection, three main processes, namely, state preparation, ansatz selection and cost function evaluation are illustrated. Further details of the VQLS algorithm can be found in Ref. [36].

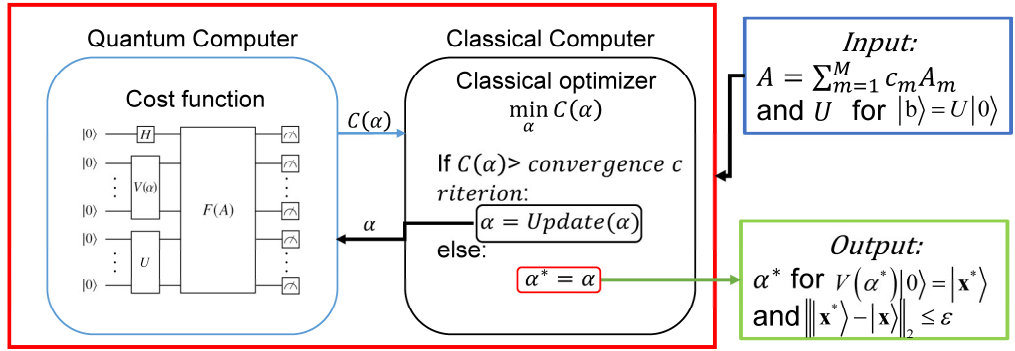

Fig. 3. Basic VQLS algorithm schematic diagram.

In the process of state preparation, the matrix $A$ is decomposed into a linear combination of $M$ unitary matrices $A_m$ with their complex coefficients $c_m$ as follows:

$$A = \sum_{m=1}^{M} c_m A_m. \qquad (24)$$

A possible decomposition is into the Pauli basis [31,36,38,40] based on the $\sigma$-matrices of the identity $I$ and Pauli gates of $X$, $Y$ and $Z$.

$$\sigma_0 = I = \begin{bmatrix} 1 & 0 \\ 0 & 1 \end{bmatrix}, \sigma_1 = X = \begin{bmatrix} 0 & 1 \\ 1 & 0 \end{bmatrix}, \sigma_2 = Y = \begin{bmatrix} 0 & -i \\ i & 0 \end{bmatrix}, \sigma_3 = Z = \begin{bmatrix} 1 & 0 \\ 0 & -1 \end{bmatrix}. \qquad (25)$$

Any $2^n \times 2^n$ matrix $A$ can be decomposed into Pauli strings with $\sigma_n = \{0,1,2,3\}^n$. Here, we denote the unitary matrix $A_m$ as one particular element of $\sigma_n$. Its corresponding coefficients $c_m$ can be determined by $c_m = \mathrm{Tr}(A_m \cdot A)/2^n$ where Tr stands for the trace of a matrix. The quantum state $|b\rangle$ can be prepared by applying a unitary operation $U$ to the ground state $|0\rangle$, which may be fulfilled by utilizing the method reported in Ref. [48].

The VQLS algorithm requires an ansatz for the gate sequence $V(\boldsymbol{\alpha})$ to simulate a potential solution $|x\rangle = V(\boldsymbol{\alpha})|0\rangle$. There are many choices of the ansatz for specific problems, and the widely used fixed structure hardware-efficient ansatz [36,49] is employed in the present study. This ansatz is comprised of multiple layers of controlled gates across alternating pairs of neighboring qubits entangled by rotation gates $R_y(\boldsymbol{\alpha})$. The structure of quantum gates is fixed and only the parameters $\boldsymbol{\alpha}$ in rotation gates $R_y$ update in each run of the quantum circuit.

For the cost function $C(\boldsymbol{\alpha})$ used in the VQLS algorithm, there are two aspects of requirements. For one thing, the value of the cost function should approach a very small value when the state $|\Phi\rangle = A|x\rangle$ is nearly proportional to $|b\rangle$. For another, the cost function may become very large when $|\Phi\rangle$ is close orthogonal to $|b\rangle$. Under these conditions, the cost function is given as follows.

$$C_{Gp}(\boldsymbol{\alpha}) = C_{Gp} = \langle x|H_P|x\rangle, \qquad (26)$$

with the Hamiltonian $H_P$ defined by

$$H_P = A^\dagger \left(I - |b\rangle\langle b|\right) A. \qquad (27)$$

To improve the accuracy of the algorithm, a normalization of the cost function is commonly applied. To be specific, dividing the cost function $C_{Gp}$ by the norm of $|\Phi\rangle$ yields

$$C_p = 1 - \frac{|\langle b|\Phi\rangle|^2}{\langle \Phi|\Phi\rangle}. \qquad (28)$$

Obviously, $\langle \Phi|\Phi\rangle$ and $|\langle b|\Phi\rangle|^2$ should be computed first for the evaluation of the cost function $C_p$. Their expectation values can be estimated by the Hadamard Test [36,40] which is a standard quantum computation technique. Finally, by minimizing $C_p$ with respect to the variational parameters, the solution $|x\rangle$ can be obtained.

It should be noted that the global expression of the cost function in Eq. (28) requires that all unitary operations ($U$, $U^\dagger$, $A^\dagger_m$, $A_m$, $V^\dagger$ and $V$) are controlled by an external ancillary qubit. This may cause experimental challenges in the Hadamard test, especially when the ansatz $V$ is comprised of many layers. Hence, the practical implementation in the present study uses a local cost function that can be measured with ease and leads towards the same optimal solution given by [36]

$$C = 1 - \frac{\sum_{m=1}^{M}\sum_{m'=1}^{M}\langle 0|V^\dagger A_{m'}^\dagger U P U^\dagger A_m V|0\rangle c_m c_{m'}^*}{\sum_{m=1}^{M}\sum_{m'=1}^{M}\langle 0|V^\dagger A_{m'}^\dagger A_m V|0\rangle c_m c_{m'}^*}, \tag{29}$$

with

$$P = \frac{1}{2}I + \frac{1}{2n}\sum_{l=0}^{n-1} Z_l, \tag{30}$$

where $Z_l$ is the Pauli Z operator locally acted on the $l$th qubit. As a result, Eq. (29) is rewritten as

$$C = \frac{1}{2} - \frac{1}{2n}\frac{\sum_{l=0}^{n-1}\sum_{m=1}^{M}\sum_{m'=1}^{M} u_{m,m',l} c_m c_{m'}^*}{\sum_{m=1}^{M}\sum_{m'=1}^{M} u_{m,m',-1} c_m c_{m'}^*}, \tag{31}$$

with the coefficients

$$u_{m,m',l} = \langle 0|V^\dagger A_{m'}^\dagger U Z_l U^\dagger A_m V|0\rangle. \tag{32}$$

The operaor $Z_l$ can be replaced with the identity if $l = -1$, i.e.,

$$u_{m,m',-1} = \langle 0|V^\dagger A_{m'}^\dagger A_m V|0\rangle. \tag{33}$$

The Hadamard test can experimentally estimate the complex coefficients $u_{m,m',l}$ and only the unitaries $A^\dagger{}_m$, $A_m$ and $Z_l$ need to be controlled. The problem can be finally resolved by minimizing the local cost function $C$.

## 3. Results and Discussion

The performance of the present VQLS-based solver is assessed by simulating the potential flow around a circular cylinder and Stokes flow in a lid-driven cavity in two-dimensions. Considering the cast linear systems introduced in Subsection 2.2, the corresponding implementation details and results with discussions are presented in this section. To be more specific, issues concerning the accuracy and time complexity (heuristic scaling) are discussed with the simulated results. All quantum simulations are implemented using the Xanadu's Pennylane open-source library [50] with a statevector simulator as a backend.

In the simulation, a hardware-efficient ansatz $V(\alpha)$ as described in Subsection 2.3 is utilized. Based on the comparison of various classical optimizers reported in Ref. [51], the gradient-descent optimizer with adaptive learning rate, first and second moment, ADAM optimizer [52,53] is selected. Considering the great effect of the initialization parameters of the ansatz on the optimization process, they are set with random values to ensure generality. The analyses of heuristic scaling solely consider converged results and each instance is run over 10 times to compute the averaged data. In the comparison hereinafter, the results obtained by the classical solver and the present VQLS-based solver are noted as "classic" and "VQLS", respectively.

### 3.1. Potential flow around a circular cylinder

In the simulation, three sets of grids with $N = M = 4$, 8 and 16 corresponding to $n = 4$, 6 and 8, respectively, are utilized to discretize the computational domain. Based on the decomposition method introduced in Subsection 2.3, when $n = 6$ qubits are used, i.e., $N = M = 8$, the matrix $A$ can be linearly decomposed to 500 items. For the case of $n = 8$ and $N = M = 16$, there are 2056 items for the linear combination of unitaries for the matrix $A$.

The quantum-classical VQLS-based solver successfully computed the solutions of stream functions, velocity magnitudes and velocity vectors. Fig. 4 and Fig. 5 depict the results for cases of $n = 6$ ($N = 8$) and $n = 8$ ($N = 16$), respectively. For comparison, the data calculated by a classical solver using the direct method are also included. Clearly, both the solution distributions and contours obtained by the classical and VQLS-based solvers achieve good agreement. This observation confirms that the present VQLS-based method is competent to resolve such a potential flow problem.

The time complexity of the present method is further evaluated for this potential flow problem. Fig. 6 and Fig. 7 plot the heuristic scaling for the dependence on $\varepsilon$ and $n$, respectively. The approximately logarithmic dependence on $1/\varepsilon$ is shown and the dependence on $n$ appears to be linear (logarithmic in $N$), which proves that the VQLS-based method may enjoy promising efficiency for resolving potential flow problems.

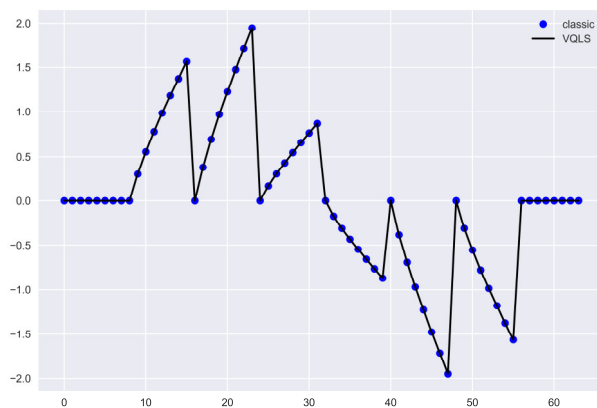

(a)

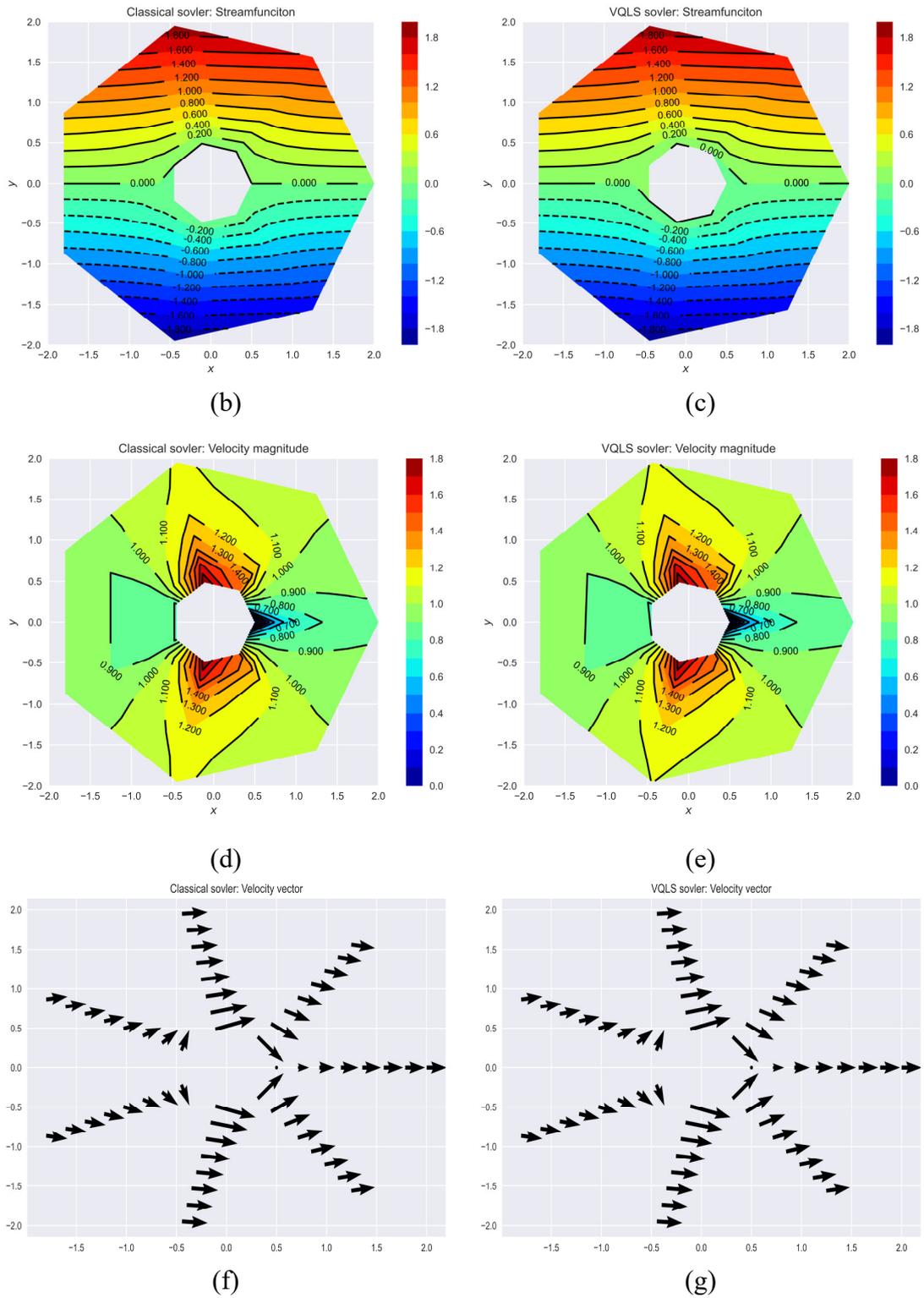

(b)                          (c)

(d)                          (e)

(f)                          (g)

Fig. 4. Results of potential flow around a circular cylinder when $N = 8$ and $\varepsilon = 0.01$: the comparison of solutions (a); the stream functions obtained by the classical solver (b) and the present VQLS-based solver (c); the velocity magnitudes obtained by the classical solver (d) and the present VQLS-based solver (e); the velocity vectors

obtained by the classical solver (f) and the present VQLS-based solver (g).

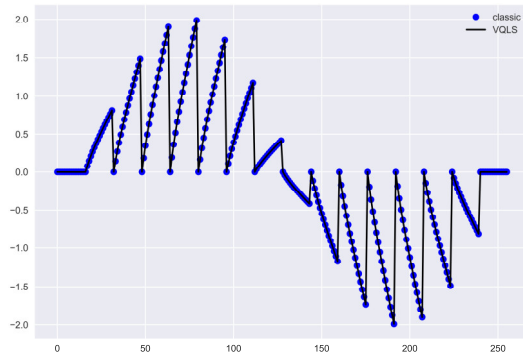

(a)

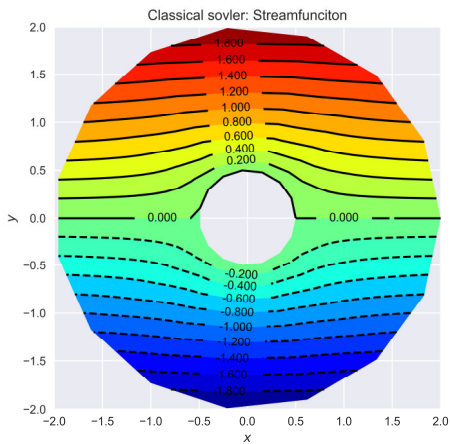 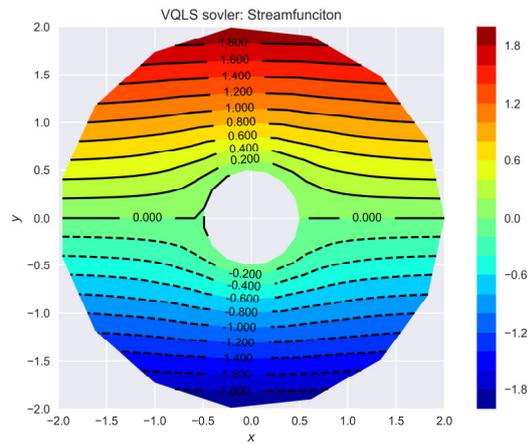

(b)　　　　　　　　　　　　　　　　(c)

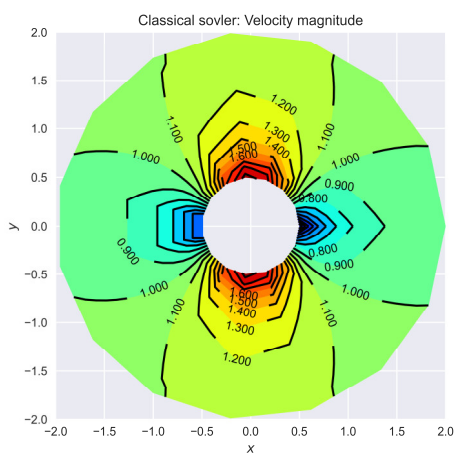 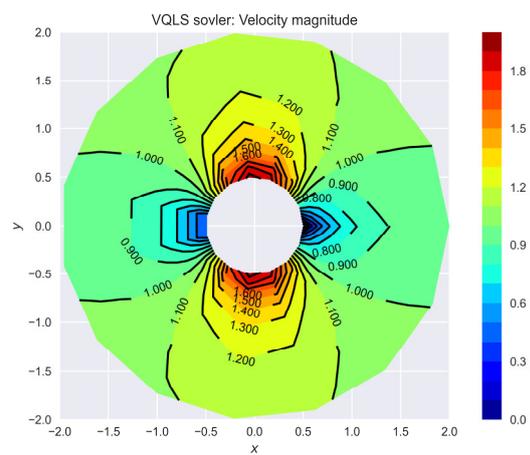

(d)　　　　　　　　　　　　　　　　(e)

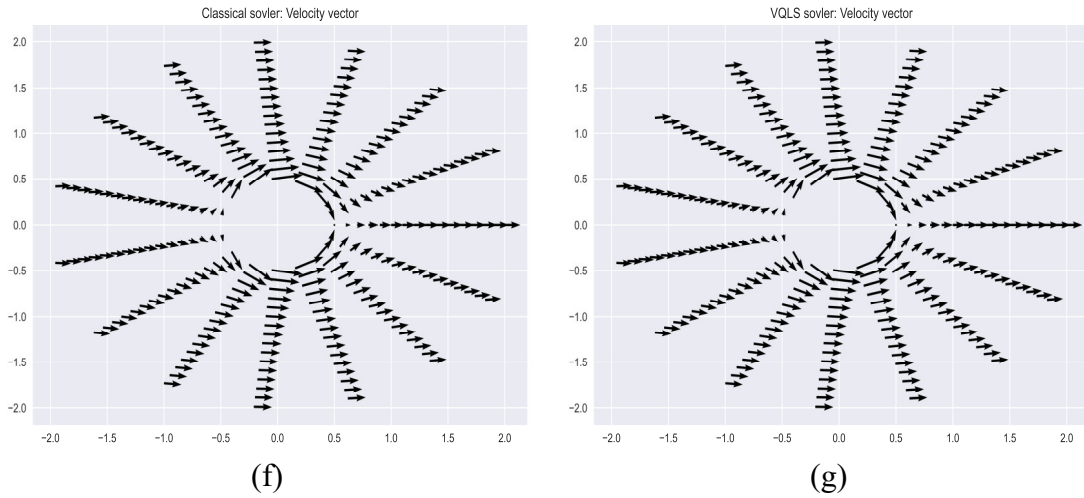

(f)                 (g)

Fig. 5. Results of potential flow around a circular cylinder when $N = 16$ and $\varepsilon = 0.01$: the comparison of solutions (a); the stream functions obtained by the classical solver (b) and the present VQLS-based solver (c); the velocity magnitudes obtained by the classical solver (d) and the present VQLS-based solver (e); the velocity vectors obtained by the classical solver (f) and the present VQLS-based solver (g).

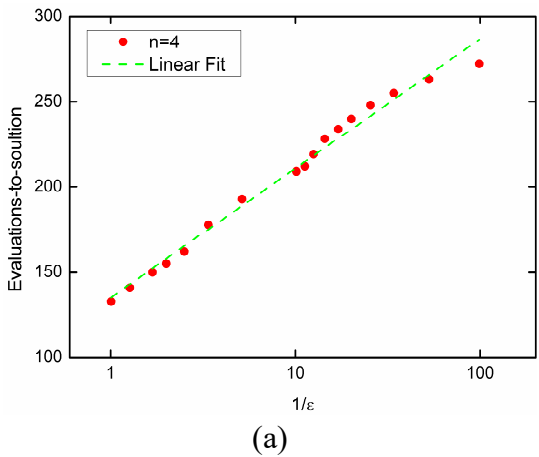 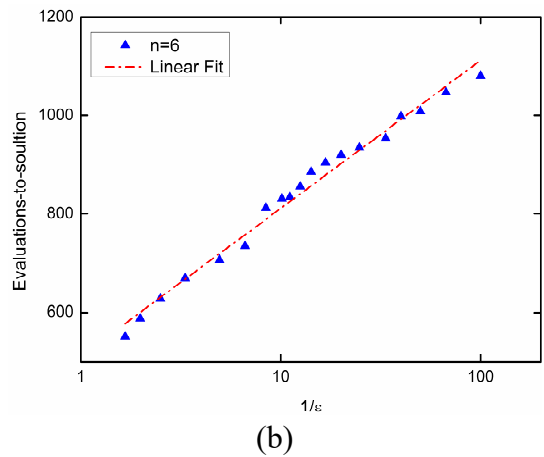

(a)                 (b)

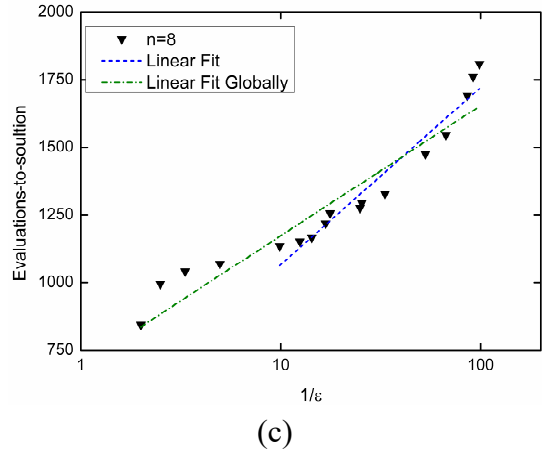

(c)

Fig. 6. Heuristic scaling for potential flow around a circular cylinder. The evaluations-to-solution versus $1/\varepsilon$ for (a) $n = 4$, (b) $n = 6$ and (c) $n = 8$. The *x*-axis is shown in the log scale.

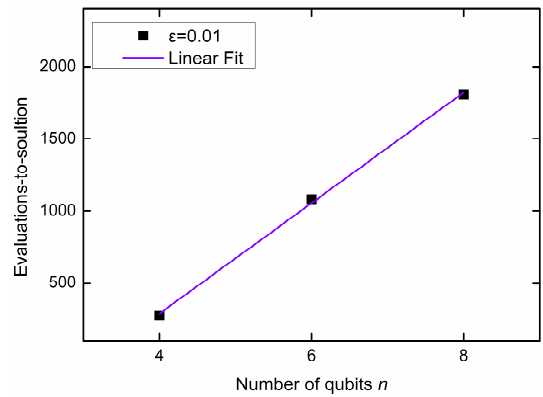

Fig. 7. Heuristic scaling with number of qubits $n$ when $\varepsilon = 0.01$ for potential flow around a circular cylinder.

### 3.2. Stokes flow in a lid-driven cavity

Since two discretization methods are applied for this Stoke flow problem, there are two sets of linear systems of equations to be solved by the VQLS algorithm. In the study of the present solver with finite difference discretization, the computations are conducted with the number of qubits $n = 4$, 6 and 8 for the number of grid points $N = 6$, 10 and 18, respectively. For all simulations, the cost function can plateau, and the results of solution distributions and contours obtained by the present VQLS-based solver agree well with those got by the classical solver. Fig. 8 and Fig. 9 plot the simulated results when $N = 10$ and 18, respectively. There is a slight difference between results of the classical and VQLS-based solvers when $N = 10$, which may be caused by the utilization of random initial parameters for the ansatz and the non-optimal optimizer. The excellent agreement shown in Fig. 9 verifies the accuracy of the present method in solving this Stokes flow problem.

For computations conducted by the GDQ discretization, the grid points in both directions are chosen as the Chebyshev nodes of the second kind. For comparison purposes, a mesh with grid points $N = 10$ is used to simulate this Stokes flow problem. By applying the previously introduced GDQ discretisation procedure, the matrix $A$ and the corresponding vector $\vec{b}$ are generated. The size of the matrix $A$ consisting of all interior grid points is $64 \times 64$ and the matrix $A$ can then be decomposed into 4042 terms. The computed results are shown in Fig. 10. Note that the solutions are fifth-order accurate (in approximating the fourth-order partial derivatives) rather than second-order accurate as in the central finite difference scheme, owing to the application of the high-order GDQ discretization. As can be seen from the comparison of the values of stream functions and the contours of stream functions in Fig. 10 (a)-(c), good agreement between the classical solver and the present VQLS-based solver is achieved. Furthermore, through interpolating the high-order GDQ solution on a refined uniform mesh of 100 × 100 points, the streamlines computed by methods using GDQ discretization are shown in Fig. 10 (d). Obviously, the high-order GDQ results presented in Fig. 10 (d) agree well with the results provided in Ref. [43]. By comparison, the second-order finite difference results given in Fig. 8 (b) and (c) exhibit much larger discrepancy with the reference data. This observation clearly demonstrates the better accuracy of the present method with GDQ discretization.

Furthermore, the time complexity is explored for this case via the heuristic scaling. Since the heuristic scaling studies may be conducted straightforwardly on the uniform mesh, only the VQLS-based method using the finite difference discretization is analyzed. Through numerical experiments with the similar scaling strategy to that in Subsection 3.1, the relationship for the evaluations-to-solution versus $1/\varepsilon$ is heuristically determined. As shown in Fig. 11 where the $x$-axis is plotted in the log scale, the data nearly can be fitted with a linear function for all values of $n$. This observation indicates that the $1/\varepsilon$ scaling is approximately logarithmic. In addition, the scaling with $n$ or $N$ guaranteeing a desired precision $\varepsilon = 0.1$ is determined and the corresponding results are presented in Fig. 12. It is found that the dependence on $n/N$ appears to be linear/logarithmic. Thus, it may be concluded that the present VQLS-based method scales efficiently in the number of qubits $n$ and the precision $\varepsilon$ for simulating the Stokes flow.

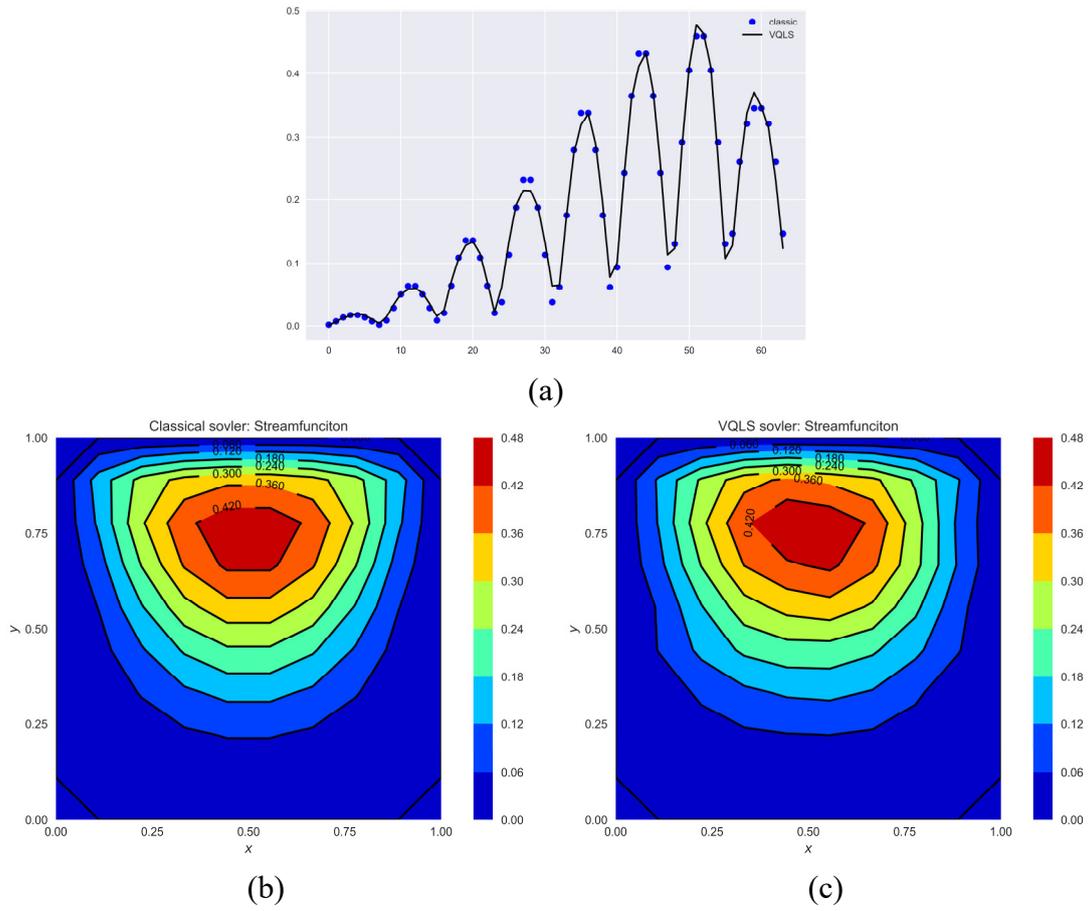

(a)

(b)                                                      (c)

Fig. 8. Results of Stokes flow in a lid-driven cavity when $N = 10$ and $\varepsilon = 0.1$: the comparison of solutions (a); the stream functions obtained by the classical solver (b) and the present VQLS-based solver (c).

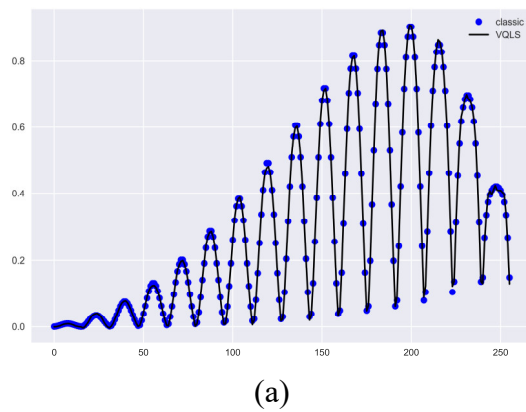

(a)

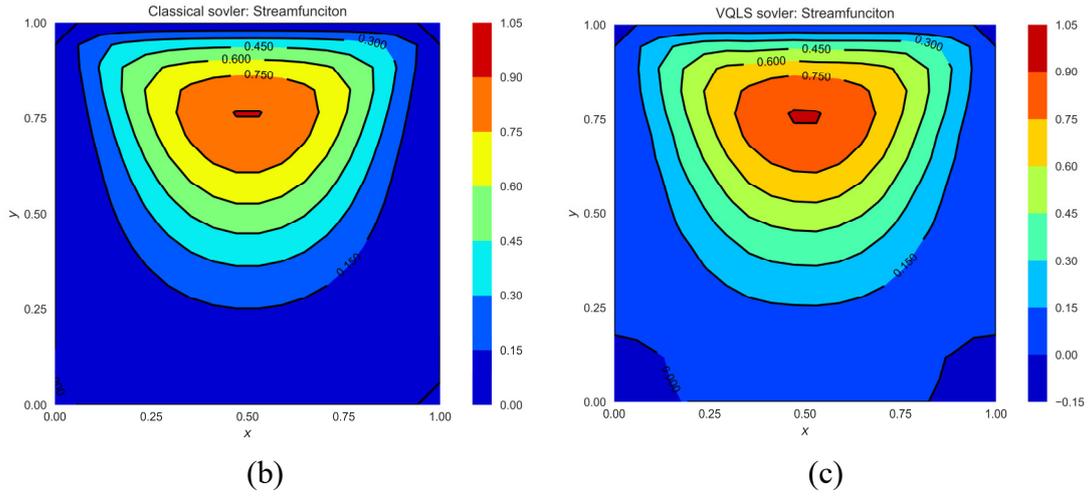

(b)                                         (c)

Fig. 9. Results of Stokes flow in a lid-driven cavity when $N = 18$ and $\varepsilon = 0.1$: the comparison of solutions (a); the stream functions obtained by the classical solver (b) and the present VQLS-based solver (c).

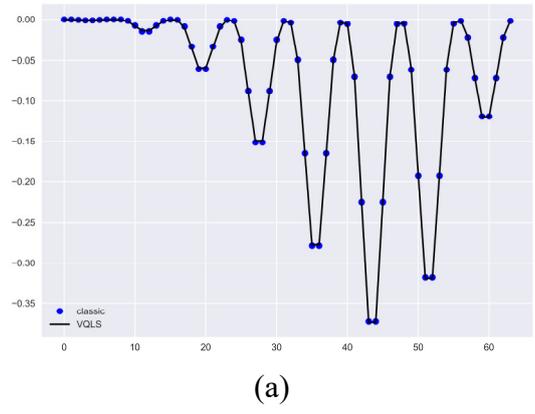

(a)

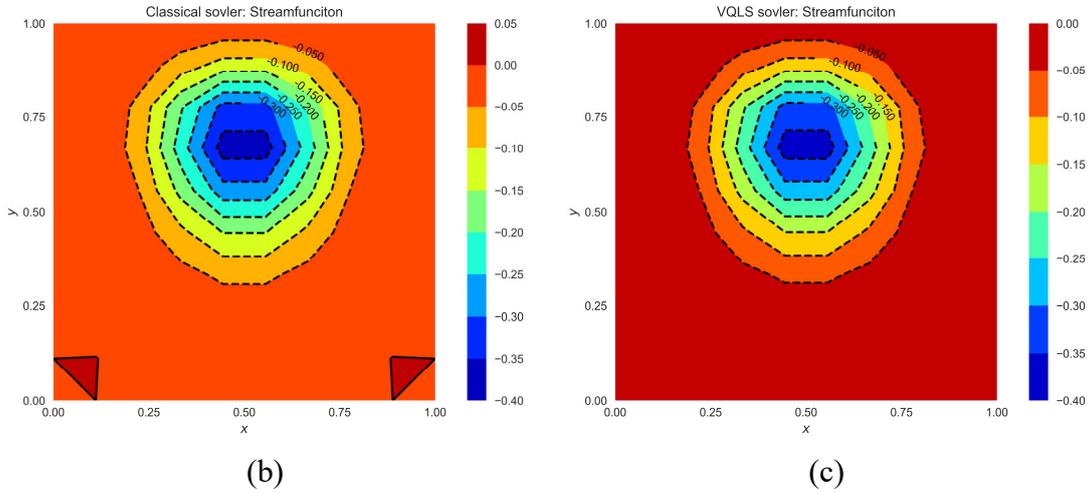

(b)                                         (c)

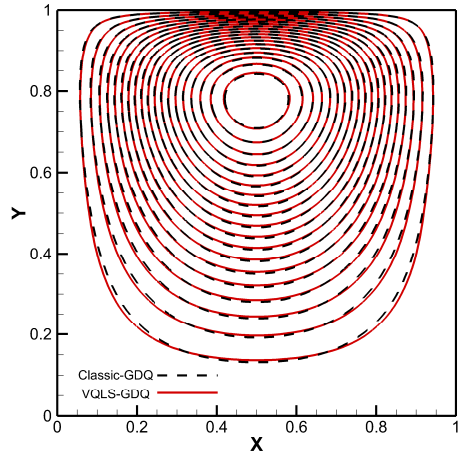

(d)

Fig. 10. Results of Stokes flow in a lid-driven cavity when $N = 10$ and $\varepsilon = 0.01$: comparison of solutions (a); stream functions obtained by the classical solver with GDQ discretization (b) and the present VQLS-based solver with GDQ discretization (c); comparison of streamlines computed by the classical solver and VQLS-based solver with GDQ discretization after post-processing.

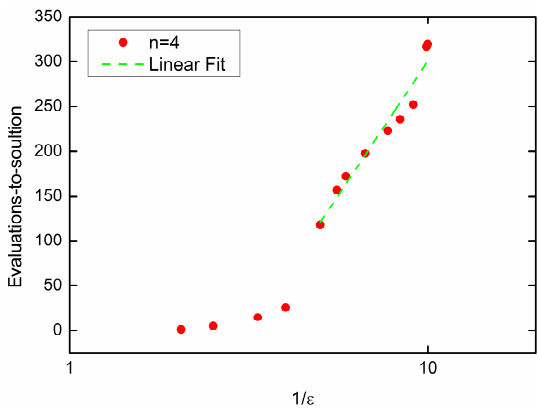
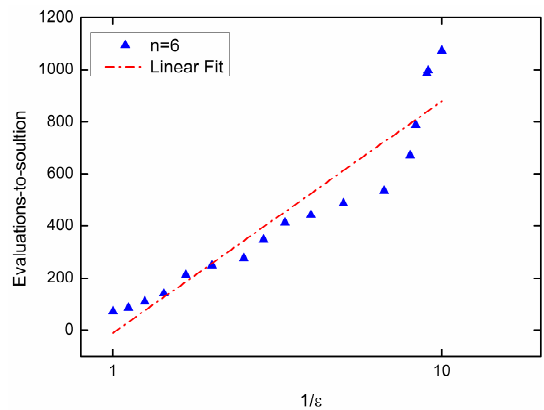

(a)                                (b)

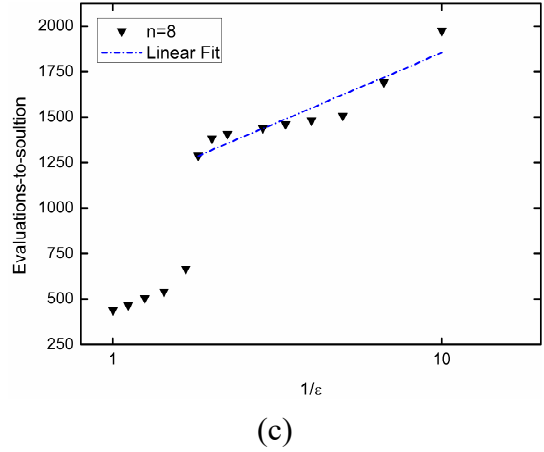

(c)

Fig. 11. Heuristic scaling for Stokes flow in a lid-driven cavity. The evaluations-to-solution versus $1/\varepsilon$ for (a) $n = 4$, (b) $n = 6$ and (c) $n = 8$. The x-axis is shown in the log scale.

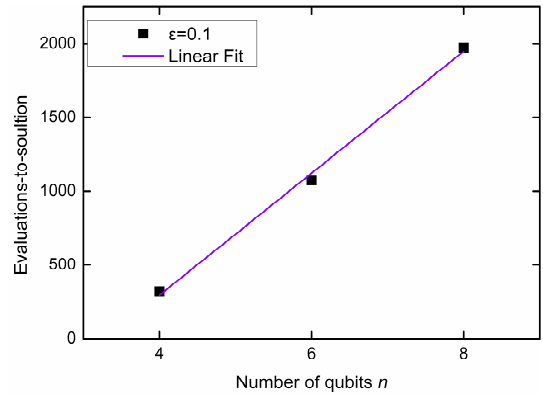

Fig. 12. Heuristic scaling with number of qubits $n$ when $\varepsilon = 0.1$ for Stokes flow in a lid-driven cavity.

## 4. Conclusions

This paper presents a direct numerical method based on the variational quantum linear solver to effectively solve governing equations of the potential flow and the Stokes flow. In this method, the governing equations are discretized by the finite difference and the generalized differential quadrature methods. Together with appropriate boundary conditions, the corresponding linear systems of equations are obtained. Note that the structure and size of the resultant linear systems from the finite difference method and the generalized differential quadrature method are different. Owing to the high-order accuracy of the GDQ method, the size of its corresponding linear systems is smaller when the same accuracy is required. Furthermore, these linear systems are solved by the variational quantum linear solver directly. In this way, the promising capability of quantum computing is brought to numerical simulation of fluid problems.

The accuracy and applicability of the present method are validated by resolving two representative problems on the quantum simulator, namely, the potential flow around a circular cylinder and Stokes flow in a lid-driven cavity. The time complexity is also examined based on the numerical results. It is demonstrated that the present quantum-classical hybrid numerical method scales efficiently in the precision and the number of qubits. Although the scale of problems simulated in this work is not very large and these problems are not extremely complex, the present method indeed provide a practical exploration and good example for incorporating the quantum computing with classical CFD solver. Based on the current good results, it is believed that the present method can be one promising alternative for efficiently solving engineering problems in the future.

**Declaration of competing interest**

The authors declare that they have no known competing financial interests or personal relationships that could have appeared to influence the work reported in this paper.